\documentstyle [12pt,axodraw] {article}

\catcode`@=12
\newcommand{\mrm}[1]{\;\mbox{\rm #1}}
\newcommand{\beq}{\begin{equation}}
\newcommand{\eeq}{\end{equation}}

\newcommand{\bea}{\begin{eqnarray}}
\newcommand{\eea}{\end{eqnarray}}

\newcommand{\Eq}[1]{Eq.~(\ref{#1})}

\newcommand{\ea}{{\it et al.}}

\newcommand{\np}[1]{{ Nucl. Phys. }{\bf #1}}
\newcommand{\plet}[1]{{ Phys. Lett. }{\bf #1}}
\newcommand{\pr}[1]{{ Phys. Rev. }{\bf #1}}

\newcommand{\zp}[1]{{ Z. Phys. }{\bf #1}}

\newcommand{\ijmp}[1]{{ Int. J. Mod. Phys. }{\bf #1}}

\def\lsim{\mathrel{\vcenter{\hbox{$<$}\nointerlineskip\hbox{$\sim$}}}}
\def\gsim{\mathrel{\vcenter{\hbox{$>$}\nointerlineskip\hbox{$\sim$}}}}

\begin{document}
\thispagestyle{empty}
\begin{flushright} UCRHEP-T233\\DESY 98-119\\August 1998\
\end{flushright}
\vspace{0.5in}
\begin{center}
{\Large \bf Probing the Exotic Particle Content\\
Beyond the Standard Model\\}
\vspace{0.7in}
{\bf Ernest Ma$^1$, Martti Raidal$^{2}$, and Utpal Sarkar$^{2,3}$\\}
\vspace{0.2in}
{$^1$ \sl Department of Physics, University of California\\}
{\sl Riverside, California 92521, USA\\}
\vspace{0.1in}
{$^2$ \sl Theory Group, DESY, D-22603 Hamburg, Germany\\}
\vspace{0.1in}
{$^3$ \sl Physical Research Laboratory, Ahmedabad 380 009, India\\}
\vspace{0.7in}
\end{center}
\begin{abstract}
We explore the possible exotic particle content beyond the standard model by 
examining all its scalar bilinear combinations.  We categorize these exotic 
scalar fields and show that without the suppression of (A) their Yukawa 
couplings with the known quarks and leptons, and (B) the trilinear couplings 
among themselves, most are already constrained to be very heavy from the 
nonobservation of proton decay and neutron-antineutron oscillations, the 
smallness of $K^0 - \overline {K^0}$, $D^0 - \overline {D^0}$ and $B_d^0 - 
\overline {B_d^0}$ mixing, as well as the requirement of a nonzero baryon 
asymmetry of the universe.  On the other hand, assumption (B) may be naturally 
violated in many models, especially in supersymmetry, hence certain exotic 
scalars are allowed to be below a few TeV in mass and would be easily 
detectable at planned future hadron colliders.  In particular, large cross 
sections for the distinctive processes like $\bar p p \to tt,\bar t c$ and 
$p p \to t t, b b$ would be expected at the Fermilab Tevatron and CERN LHC, 
respectively.
\end{abstract}

\newpage
\baselineskip 24pt

\subsection*{1. Introduction}

The quarks and leptons of the minimal standard model are familiar fixtures 
of particle physics.  Under the standard $SU(3)_C \times SU(2)_L \times 
U(1)_Y$ gauge group, they transform as follows
\begin{equation}
\left( \begin{array} {c} u_i \\ d_i \end{array} \right)_L \sim (3,2,1/6), 
~~~u_{iR} \sim (3,1,2/3), ~~~d_{iR} \sim (3,1,-1/3);
\end{equation}
\begin{equation}
\left( \begin{array} {c} \nu_i \\ l_i \end{array} \right)_L \sim (1,2,-1/2), 
~~~l_{iR} \sim (1,1,-1).
\end{equation}
In the above, the index $i$ denotes the known 3 families.  Only one scalar 
bilinear combination of these fermions is required in the minimal model, 
{\it i.e.}
\begin{equation}
(\phi^+, \phi^0) \sim (1,2,1/2),
\end{equation}
which couples $\overline {(u_i,d_i)}_L$ to $u_{jR}$ and $d_{jR}$, as well as 
$\overline {(\nu_i,l_i)}_L$ to $l_{jR}$.  As $\phi^0$ acquires a nonzero 
vacuum expectation value, $v \equiv \langle \phi^0 \rangle$, the quarks and 
leptons obtain masses and there is mixing among the quark families, but not 
among the lepton families.  
The other possible bilinear combinations may give rise to unobserved 
phenomenology. Therefore their masses and 
couplings to fermions have been stringently constrained from 
low energy and collider processes \cite{5,leptoq}.

However, the scalar bilinear combination 
which couples $(\nu_i,l_i)_L$ to $(\nu_j,l_j)_L$, 
{\it i.e.},
\begin{equation}
(\xi^{++}, \xi^+, \xi^0) \sim (1,3,1)\,,
\end{equation}
has been shown to have important phenomenological implications \cite{1}
beyond these considered in the previous works. The new ingredient is 
the  trilinear coupling of $\xi$ to the  standard model Higgs boson
which, together with the Yukawa couplings to  fermions,
may give rise to baryogenesis or, alternatively,  to wash away
the baryon asymmetry of the universe. Also, 
for a very large $M_\xi$, it is natural for $\xi^0$ to acquire a tiny vacuum 
expectation value, thereby allowing neutrinos to obtain very small Majorana 
masses and to mix with one another.

In this paper, we extend the above consideration to all possible scalar 
bilinear combinations of the quarks and leptons of the minimal standard 
model including diquark scalars.  
We categorize these exotic scalar fields and ascertain their various 
contributions to physics within and beyond the standard model. From a general 
consideration we can show that most are already constrained by present 
experimental data to be very heavy.  This is based on two assumptions: (A) 
the Yukawa couplings of these exotic scalars with the known quarks and leptons 
are all of order unity, and (B) the trilinear couplings of these exotic 
scalars among themselves are of order the mass scale of the heaviest particle 
involved.  Whereas assumption (A) is a natural one in almost any model, 
assumption (B) is subject to many other possible qualifications.  For example, 
exact supersymmetry often forbids the existence of trilinear scalar 
couplings, in which case any such coupling should not exceed the scale of 
supersymmetry 
breaking, which may be very small compared to the mass of the heaviest 
particle involved.  Relaxing assumption (B) allows certain exotic scalar 
particles to be below a few TeV in mass and they would easily be detectable at 
planned future hadron colliders.  In particular, large cross sections for the 
flavor-changing neutral-current (FCNC) process $\bar p p \to \bar t c$, and 
for the quark flavour violating resonance processes $p p \to t t, b b$ 
may be expected at the Fermilab Tevatron and CERN LHC, respectively.
Also, the cross section of the resonance 
process $\bar p p \to  t t$ at Tevatron, 
though sea quark suppressed, may be large enough to provide an observable
excess of the same-sign dilepton final states indicating clearly for
the new physics. 

The outline of the paper is as follows. After the Introduction
we classify the exotic scalar bilinears and discuss constraints
on their masses and couplings. In Section 3 we present our results on
diquark mediated processes at hadron colliders. We conclude in Section 4.

\begin{table}
\begin{center}
\begin{tabular}{|c|c|c|c|c|c|}
\hline
Representation & $qq$ & $\bar q \bar l$ & $q \bar l$ & $ll$ & Comment \\
\hline
$(1,1,-1)$ & & & & X & $n_f \geq 2$ \\
$(1,3,-1)$ & & & & X & neutrino masses \\
$(1,1,-2)$ & & & & X & $e^- e^-$ collider \\
\hline
$(3^*,1,1/3)$ & X & X & & & $p \to e^+ \pi^0$ \\
$(3^*,3,1/3)$ & X & X & & & $n_f \geq 2$ \\
$(3^*,1,4/3)$ & X & X & & & $n_f \geq 2$ \\
$(3^*,1,-2/3)$ & X & & & & $n_f \geq 2$ \\
\hline
$(3,2,1/6)$ & & & X & & HERA \\
$(3,2,7/6)$ & & & X & & anomaly \\
\hline
$(6,1,-2/3)$ & X & & & & neutron- \\
$(6,1,1/3)$ & X & & & & antineutron \\
$(6,1,4/3)$ & X & & & & oscillations \\
\hline
$(6,3,1/3)$ & X & & & & $K^0 - \overline {K^0}$ and  \\
$(8,2,1/2)$ & & & & & $D^0 - \overline {D^0}$ mixing \\
\hline
\end{tabular}
\caption{Exotic scalar particles beyond the standard model.}
\end{center}
\end{table}

\subsection*{2. Classification and Indirect Constraints}

In this section we classify the exotic scalars and constrain their masses and 
couplings from low-energy phenomenology and from cosmological considerations. 
In Table 1 we list all possible scalar representations which are products of 
two known fermion representations.  
The colour singlet fields have been listed before in \cite{5} and the
colour triplets and sextets in \cite{volkas}.
We cite the most stringent constraints on their masses and interactions
from previous works and add several new bounds. 

\subsubsection*{2.1. Dileptons}

First we consider the three scalar dileptons.

{\bf (1,1,--1):} 
This scalar singlet couples to $\nu_i l_j - l_i \nu_j$ where $i 
\neq j$, hence the number of families $n_f$ must be 2 or greater.  
It contributes to $\mu$ and $\tau$ decays 
leading to the constraints
 \cite{5,4}
\begin {eqnarray}
\mu \to e \gamma &:& {M_X \over (f_{e \tau} f_{\mu \tau})^{1/2}} 
 \gsim 1.6 \times 10^4 \mrm{GeV}\,; \nonumber \\
G_\tau &:& {M_X \over f_{e \tau}} \gsim 2.2 \times 10^3 \mrm{GeV} , ~~~
{M_X \over f_{\mu \tau}} \gsim 3.2 \times 10^3 \mrm{GeV}\,;
\end{eqnarray}
where $M_X$ denotes the scalar mass.
Similar bounds can be derived 
from the precision electroweak data. 
If it is assumed that this exotic scalar contributes less than 0.1\% of the 
$\mu$ decay rate, then  its mass divided by 
$|f_{e \mu}|$ is greater than $1.1 \times 10^4$ GeV \cite{3}.
It has been also shown to be  
useful for the radiative generation of neutrino masses \cite{2}.

In models with two or more Higgs doublets, there are in general trilinear 
couplings given by
\bea  
{\cal L} = h_1 X^+ (\nu_i l_j - l_i \nu_j) + h_2 M_X X^- 
(\phi_i^+ \phi_j^0 - \phi_i^0 \phi_j^+)\,. 
\eea
Lepton-number conservation is thus violated and if $h_1$ and $h_2$ are 
unsuppressed (which is the case in most models of radiative neutrino masses), 
the resulting interactions will erase any lepton asymmetry of the universe 
before the onset of the electroweak phase transition.  This will deprive 
the anomalous sphaleron-induced processes from converting an existing 
lepton asymmetry into a baryon asymmetry of the universe \cite{lepto}.  
To forestall this eventuality, $X$ must be heavy: 
\bea
{M_X \over h_1^2} \gsim 10^{15} \mrm{GeV} ~~~{\rm or} ~~~ 
{M_X \over h_2^2} 
\gsim 10^{15} \mrm{GeV} ~~~ {\rm and} ~~~{M_X \over h_1^2 h_2^2} \gsim 10^{16} 
\mrm{GeV}. 
\eea

{\bf (1,3,--1):} 
This scalar triplet couples according to $\xi^0 \nu_i \nu_j + 
\xi^+ (\nu_i l_j + l_i \nu_j)/\sqrt 2 + \xi^{++} l_i l_j$ and allows us 
to have the most general $3 \times 3$ Majorana neutrino mass matrix 
without right-handed neutrinos.  It also couples to the standard Higgs 
doublet according to $\bar \xi^0 \phi^0 \phi^0 + \sqrt 2 \xi^- \phi^+ \phi^0 
+ \xi^{--} \phi^+ \phi^+$ and it has been shown \cite{1} that a tiny 
$\langle \xi^0 \rangle$ of a few eV or less is obtained for 
$M_\xi\gsim 10^{13}$ GeV.  If there are two such triplets, a 
successful leptogenesis scenario \cite{1} for the baryon asymmetry of the 
universe may also be obtained.

Alternatively, it is possible that $\xi$ does not couple to $\phi$, in which 
case lepton number is conserved, but not lepton flavor.  In this latter 
scenario, the most stringent 
constraints come from $\mu \to e e \bar{e}$ \cite{5} decay and
$\mu - e$ conversion in nuclei \cite{RS},  
which are given by 
\bea
{M_X \over (f_{ee} f_{e \mu})^{1/2}} \gsim 2.2 \times 10^5 \mrm{GeV}
~~~{\rm and}~~~ 
{M_X \over (f_{e a} f_{\mu a})^{1/2}} \gsim 4 \times 10^4 \mrm{GeV} \,,
\eea
respectively,
where, $a = e, \mu $ or $\tau$.  The doubly charged component of the triplet 
may give rise to unique lepton-flavor-violating signatures at collider 
experiments \cite{5,double,R}.  In particular, for certain neutrino mass 
patterns where the neutrino decay lifetimes are constrained, there are lower 
bounds on the cross sections of the processes $e^-e^- (\mu^-\mu^-)\to l^-l^-$ 
such that they {\it must} be seen at future lepton colliders \cite{R}.

{\bf (1,1,--2):} This doubly charged scalar singlet couples to $l_{iR} l_{jR}$ 
symmetrically.  It is relevant for $e^- e^-$ colliders \cite{5,double}, but 
it also contributes to lepton-flavor-changing processes such as $\mu-e$ 
conversion, $\mu \to e \gamma,$ and $\mu \to e e e$.  The bounds from these 
processes are the same as those for the (1,3,--1) scalar.

If this exotic scalar $X$ coexists with the previously discussed $(1,1,-1)$, 
now call it $Y$, which is much lighter than $X$, then there may be 
lepton-number-violating couplings given by
$$ {\cal L} = h_1 X^{++} l_R l_R + h_2 M_X X^{++} Y^- Y^- . $$
In this case, $X$ has to be again heavy to satisfy
the leptogenesis constraints \cite{lepto},
\bea
 {M_X \over h_1^2} \gsim 10^{15} \mrm{GeV} ~~~{\rm or} ~~~ {M_X \over h_2^2} 
\gsim 10^{15} \mrm{GeV} ~~~ {\rm and} ~~~{M_X \over h_1^2 h_2^2} \gsim 10^{16}
 \mrm{GeV}. 
\eea

\subsubsection*{2.2. Triplet Diquarks and Leptoquarks}

The decomposition of $3 \times 3$ being $3^* + 6$ under SU(3), there are two 
types of scalar diquarks.  The antisymmetric combination $3^*$ may also 
couple to an antiquark and an antilepton as shown in Table 1.  This means 
that proton decay is always possible, if not at tree level \cite{volkas}
 then in one loop as we show below.

{\bf (3*,1,1/3):}  This exotic scalar singlet couples to $u^\alpha_{iL} 
d^\beta_{jL} - u^\beta_{iL} d^\alpha_{jL} - d^\alpha_{iL} u^\beta_{jL} + 
d^\beta_{iL} u^\alpha_{jL}$ where $\alpha, \beta = 1,2,3$ 
are color indices.  This combination is antisymmetric under both $SU(3)_C$ 
and $SU(2)_L$, but symmetric under the interchange of families.  It also 
couples to $u^\alpha_{iR} d^\beta_{jR} - u^\beta_{iR} d^\alpha_{jR}$ as well 
as $\bar u_{iL} \bar l_{jL} - \bar d_{iL} \bar \nu_{jL}$ and $\bar u_{iR} 
\bar l_{jR}$.  Hence it mediates proton decay with the effective operators 
($u u d l$) and ($u d d \nu$).  Using $\tau (p \to \pi^0 e^+) \gsim 9 \times 
10^{32}$ years, one finds the constraint
\bea
 \frac{M_X}{|f_{ud} f_{ul}|^{1/2}}\gsim 10^{16} \mrm{GeV.} 
\eea
Although this is the strongest bound, it is applicable only to $u$, $d$, 
and $s$ quarks.  However, if other couplings are large, there will be fast 
baryon-number-violating interactions in the early universe and a baryon 
asymmetry cannot be maintained. Thus strong bounds exist for all quarks, 
namely 
\bea
\frac{M_X}{f^2} \gsim 10^{15} \mrm{GeV.}  
\eea
On the other hand, if baryon asymmetry 
is generated after these scalars have decayed away, then these bounds will 
not be valid.

{\bf (3*,3,1/3):}  
This triplet scalar diquark couples to a symmetric combination 
of $SU(2)_L$ doublets, hence it must be antisymmetric in its coupling to 
quark families. It mediates proton decay with the effective operators 
($u u s l$) and ($u d s \nu$).  From the nonobservation \cite{4} of processes 
such as $p \to e^+ K^0$ or $p \to \mu^+ K^0$, we find its mass divided by 
$|f_{us} f_{ul}|^{1/2}$ to be also greater than about 10$^{16}$ GeV.

There will also be constraints from the survival of the baryon
asymmetry of the universe in this case,
which are of similar magnitude but applicable to all generations, 
$M_X/ f^2 \gsim 10^{15}$ GeV.

{\bf (3*,1,4/3):}  As a diquark, this exotic scalar couples only to $u_{iR} 
u_{jR}$,  where $i \neq j$.  As a leptoquark, it couples only to $\bar d_{iR} 
\bar l_{jR}$.  Since the $c$ or $t$ quark must appear in a tree-level 
effective operator, proton decay here requires a one-loop diagram such 
as the one depicted in Fig.~1.  The resulting effective operator ($u u s l$) 
is suppressed by the factor $G_F m_c m_s V_{cs} V_{us} / 16 \pi^2$.  We 
find thus the constraint on the mass of this scalar to be
\bea
\frac{M_X}{|f_{uc} f_{sl}|^{1/2}}\gsim 5.3 \times 10^{11} \mrm{GeV.}
\eea
There will also be constraints from baryogenesis in this case,
which are of similar magnitude but applicable to all generations, 
$M_X/ f^2 \gsim 10^{15}$ GeV. 

{\bf (3*,1,--2/3):}  
Because there is no singlet neutrino, this exotic scalar 
acts as a diquark, but not a leptoquark.  It couples only to $d_{iR} d_{jR}$, 
where $i \neq j$.  Hence it does not appear to mediate proton decay. 
However, the trilinear scalar interaction
\begin{equation}
(3^*,1,-2/3) (3^*,1, 1/3) (3^*,1,1/3)
\end{equation}
is generally allowed and the effective operators $(u d s \bar \nu)$ and 
$(d d s \bar l)$ may be obtained in one loop as depicted in Fig.~2.  Note 
that these have the selection rule $\Delta B = -\Delta L$ instead of the 
usual $\Delta B = \Delta L$.  Let the trilinear coupling be $\mu$ and the mass 
of the $(3^*,1,1/3)$ scalar be $M_Y$, then the loop suppression factor for 
decays such as $n \to e^- K^+$ and $p \to \nu K^+$ is $\mu m_p / 16 \pi^2 
M_Y^2$, where $m_p$ is the proton mass.  Taking the natural value of $\mu$ 
to be equal to $M_Y \simeq 10^{16}$ GeV, we find the mass of the 
$(3^*,1,-2/3)$ scalar divided by $|f_{ds} f_{ud} f_{d\nu}|^{1/2}$ to be 
greater than about 
\bea
\frac{M_X}{|f_{ds} f_{ud} f_{d\nu}|^{1/2}}
\gsim \frac{(m_p M_Y)^{1/2}}{4\pi} \simeq 7.7 \times 10^6 \mrm{GeV.}
\eea
In addition to this bound, if the scalar (3*,1,--2/3) has a trilinear 
coupling as assumed above, then its interaction with two quarks and 
simultaneously with two (3,1,--1/3) scalars will break baryon number. 
To prevent an existing baryon asymmetry from being washed out, we would 
then get a much stronger bound on its mass and couplings, 
\bea
M_X^3/\mu^2\gsim 10^{15} \mrm{GeV} \;\;\;\mrm{or}\;\;\; 
M_X/f^2 \gsim 10^{15} \mrm{GeV} \;\;\;\mrm{and}\;\;\; 
M_X^3/f^2 \mu^2 \gsim 10^{16} \mrm{GeV,} 
\eea
assuming that $M_Y \ll M_X$.  The same consideration 
applies to the (3*,1,4/3) scalar if the trilinear coupling (3*,1,4/3) 
(3*,1,--2/3) (3*,1,--2/3) also exists.

\subsubsection*{2.3. Leptoquarks}

There are two scalar leptoquark representations, both of which are $SU(2)_L$ 
doublets.  They are $q \bar l$ (and not $q l$) combinations, and may be 
relevant \cite{9} to the erstwhile HERA anomaly \cite{10} of excess events 
at large momenta in $e^+ p$ scattering. 
Reviews on the constraints on leptoquark couplings to fermions
can be found in \cite{leptoq}.

{\bf (3,2,1/6):}  This scalar leptoquark couples to $(\bar l_i, \bar \nu_i)_L 
d_{jR}$.  It mediates lepton flavor-changing processes such as $K^0 \to 
e^+ \mu^-$.  From the experimental upper bound \cite{4} of the latter, one 
finds its mass divided by $|f_{de} f_{s\mu}|^{1/2}$ to be greater than 
about $2.4 \times 10^5$ GeV. 

{\bf (3,2,7/6):}  
This scalar leptoquark couples to $(\bar l_i, \bar \nu_i)_L 
u_{jR}$ and $\bar l_{iR} (u_j, d_j)_L$.  The same constraint applies here 
as in the previous case.

\subsubsection*{2.4. Sextet Diquarks}

The other type of scalar diquark is an $SU(3)_C$ sextet.  It couples 
symmetrically to two $SU(3)_C$ quark triplets.  It has not received much 
attention in the past, but it is potentially an important hint for physics 
beyond the standard model. These scalars contribute at tree level to 
 $K^0 - \overline {K^0},$  $D^0 - \overline {D^0}$ and 
 $B_d^0 - \overline {B_d^0}$ mixings, and 
 neutron-antineutron oscillations naturally 
occur from the trilinear scalar interactions $(6,1,-2/3)(6,1,1/3)^2$, 
$(6,1,-2/3)(6,3,1/3)^2$ and $(6,1,4/3)(6,1,-2/3)^2$.

{\bf (6,1,--2/3):}  This exotic scalar couples to $d_{iR} d_{jR}$ 
symmetrically, whereas $(3^*,1,-2/3)$ does so antisymmetrically as already 
discussed.  Comparing the two cases, we see that a similar loop diagram to 
Fig.~2 would generate the effective operators $(u d d \bar \nu)$ and 
$(d d d \bar l)$.  Hence its mass divided by $|f_{dd} f_{ud} f_{d\nu}|^{1/2}$ 
should also be greater than about $7.7 \times 10^6$ GeV.
The effective $dd\to ss$ and $dd\to bb$ transitions which induce   
$K^0 - \overline {K^0}$ and $B_d^0 - \overline {B_d^0}$
mixings give  somewhat weaker bounds, $1.5\times 10^6$ GeV and 
$4.6\times 10^5$ GeV, on its mass over the couplings 
$ |f_{dd}f_{ss}|^{1/2}$ and $ |f_{dd}f_{bb}|^{1/2},$ respectively.

{\bf (6,1,1/3):}  This exotic scalar couples to $u^\alpha_{iL} d^\beta_{jL} + 
u^\beta_{iL} d^\alpha_{jL} - d^\alpha_{iL} u^\beta_{jL} - d^\beta_{iL} 
u^\alpha_{jL}$, which is antisymmetric under $SU(2)_L$ and the interchange of 
families.  It also couples to $u^\alpha_{iR} d^\beta_{jR} + u^\beta_{iR} 
d^\alpha_{jR}$.  Since the trilinear scalar interaction
\begin{equation}
(6,1,1/3) (6,1,-2/3) (6,1,1/3)
\end{equation}
is generally allowed, the analog of Fig.~2 is also possible here.  We find 
the mass of the $(6,1,1/3)$ scalar divided by $|f_{ud} f_{db} 
f_{b\nu}|^{1/2}$ to be also greater than about $7.7 \times 10^6$ GeV.

As discussed before, the coexistence of a trilinear scalar coupling and a 
Yukawa coupling with the quarks will cause fast baryon-number violation. 
Forbidding it will impose
a stronger bound on the mass and couplings of the heavier scalar, 
$M_X/f^2 \gsim 10^{15}$ GeV. 

{\bf (6,1,4/3):}  
This exotic scalar couples to $u_{iR} u_{jR}$ symmetrically. 
It does not participate in proton decay.  However, from the allowed 
trilinear scalar interaction
\begin{equation}
(6,1,4/3) (6,1,-2/3) (6,1,-2/3),
\end{equation}
we obtain the effective $(udd)^2$ operator which induces neutron-antineutron 
oscillations.  From the present experimental upper bound \cite{11} of the 
latter, we find the mass of the $(6,1,4/3)$ scalar divided by $|f_{dd}|
|f_{uu}|^{1/2}$ to be greater than about $1.3 \times 10^3$ GeV.  However, 
this limit is superceded by considering the effective $uu \to cc$ transition 
which induces $D^0 - \overline {D^0}$ mixing.  From the experimental upper 
bound \cite{4} $\Delta m_D < 1.4 \times 10^{-10}$ MeV, we find the mass of 
the $(6,1,4/3)$ scalar divided by $|f_{uu} f_{cc}|^{1/2}$ to be greater than 
about $7.3 \times 10^5$ GeV.  Again the consideration of baryogenesis applies 
as before.

{\bf (6,3,1/3):}  This exotic $SU(2)_L$ triplet has all the couplings of the 
previous 3 singlets.  It partcipates in the trilinear scalar interaction
\begin{equation}
(6,3,1/3) (6,3,1/3) (6,1,-2/3),
\end{equation}
which induces $n-\bar n$ oscillations.  Hence its mass divided by 
$|f_{ud}|^{1/2}|f_{dd}|^{1/4}$ should be greater than about $1.0 \times 10^5$ 
GeV.  The baryogenesis bound will also be similar as in the other cases.
Previously quoted bounds coming from the 
 $K^0 - \overline {K^0},$
$D^0 - \overline {D^0}$ and $B_d^0 - \overline {B_d^0}$ mixings 
apply also in this case..

\subsubsection*{2.5. Octet}

{\bf (8,2,1/2):}  This $SU(3)_C$ octet couples $(\overline {u^\alpha_i, 
d^\alpha_i})_L$ to $u^\beta_{jR}$ and $d^\beta_{jR}$.  It carries neither 
baryon nor lepton number.  Consider the neutral member of this $SU(2)_L$ 
doublet.  Unlike the standard neutral Higgs boson, it has in general 
nondiagonal couplings to quarks.  Hence there is an effective $d \bar s \to 
s \bar d$ transition which induces $K^0 - \overline {K^0}$ mixing.  We find 
thus the mass of this scalar divided by $|f_{ds}f_{sd}|^{1/2}$ 
to be greater than about $1.5 \times 10^6$ GeV. Bounds from the other 
neutral meson mixings are also valid here.

All the bounds coming from the survival of the baryon asymmetry of the
universe may be avoided to some extent, if a baryon asymmetry
of the universe is generated after the heavy scalars (whose interactions 
violate baryon number) have all decayed away.  However, in this case, there 
will still be a bound on the masses of these heavy scalars, which is the 
scale of baryogenesis.  If these scalars are lighter than the scale at 
which baryon asymmetry of the universe is generated, then their 
interactions will erase the asymmetry thus generated.  Hence the bounds 
from the constraints of baryogenesis can at most be made milder by 
generating a baryon asymmetry of the universe at a lower energy scale. 

\subsection*{3. Diquarks at Hadron Colliders}

In general,
the above derived bounds on the couplings and masses of the $SU(3)_C$
triplet and sextet diquarks depend on the attributed flavor indices 
and, most importantly, on the assumption of the existence of 
the trilinear couplings.  For example, if the latter are absent or strongly 
suppressed as in some models of supersymmetry, then the most stringent bounds from
proton decay, neutron-antineutron oscillations and baryogenesis can be 
evaded in the case of sextet diquarks.  The bounds coming from the 
measurements of $K^0 - \overline {K^0}$,  $D^0 - \overline {D^0}$
and $B^0_d - \overline {B^0_d}$  mixings still apply, but if we take care 
of them by suppressing the diquark diagonal couplings involving the first 
two families, then no bounds at all exist for the others. The diquark 
couplings to the third family, in particular to the top quark, 
can only be tested in experiments at  Tevatron and LHC.
In order to be consistent with the above discussed experimental constraints 
there are widespread arguments \cite{fc} that new flavor-changing 
interactions are likely to affect mainly the physics of the third 
family of quarks only. 

At  Tevatron  diquarks may give rise to $t$-channel processes
$u\bar u\to \bar u_i u_j,$ $u\bar d\to \bar d_i u_j$ and 
$d\bar d\to \bar d_i d_j$ due to the
proton and anti-proton valence quarks collisions; and  
$s$-channel resonance processes 
$u u\to  u_i u_j,$ $u d\to  d_i u_j$ and $d d\to  d_i d_j$ 
due to the proton valence quarks and anti-proton sea quarks collisions.
Among these the most interesting 
final states are $t\bar t,$ $t t,$ $t\bar b,$ $t b,$ $b\bar b,$ $bb$ 
and $t \bar c,$ $t  c$ because top, bottom and
charm tagging allows one to distinguish them effectively from the 
standard model background.
In particular, the observation of a large rate for the flavor-changing 
process $uu\to t\bar c$ at the Tevatron Runs II and III would indicate
unambiguously the existence of new physics \cite{eilam}.
Also, large excess of the same-sign dileptons from the $tt$ final states
may indicate for the new physics.

Because collider phenomenology of the diquarks depends on their quantum
numbers we have to discuss their couplings in more detail.
The form of the Yukawa Lagrangian describing the $SU(2)_L$ singlet 
diquark $X$ coupling to down type quarks is given by
\bea
{\cal L}=f_{ij}\, \left[T_{ab}\right]^m
\,\overline{(d^c_{R})^a_i}\; d^b_{Rj}\, X_m + h.c. \,,
\label{lag1}
\eea
where $\overline{d^c}=-d^T C^T,$ 
$i,j=1,2,3$ are family indices, $a,b=1,2,3$ are colour indices,
 $R$ denotes the chirality of the quarks
and matrices $T$ form the basis of the $n$ dimensional $3\times 3$ matrix 
representation of the 
$SU(3)_c$ group. Therefore $m=1,...,n.$
For colour triplets ($n=3$) the matrices $T_{ab}$ should be anti-symmetric in
$a$ and $b$ and one may identify  $[T_{ab}]^m=\varepsilon_{ab}^{\;\;\;\;m},$
where $\varepsilon_{ab}^{\;\;\;\;m}$ is the 
three dimensional totally anti-symmetric tensor.
For colour sextets ($n=6$) the matrices $T_{ab}$ are symmetric in
$a$ and $b.$ 
Analogously to \Eq{lag1} we have for the $SU(2)_L$ singlet diquark 
couplings to up and down type quarks 
\bea
{\cal L}=f_{ij}\, \left[T_{ab}\right]^m
\,\overline{(u^c_{R})^a_i}\; d^b_{Rj}\, X_m + h.c. \,,
\label{lag2}
\eea
and for the  couplings to up  type quarks
\bea
{\cal L}=f_{ij}\, \left[T_{ab}\right]^m
\,\overline{(u^c_{R})^a_i}\; u^b_{Rj}\, X_m + h.c. \,.
\label{lag3}
\eea
Yukawa interactions of the $SU(2)_L$ triplet diquarks to quarks 
can be expressed by the Lagrangian
\bea
{\cal L}=f_{ij}\, \left[T_{ab}\right]^m\,
(\tau_\kappa\varepsilon)^{\alpha\beta}
\,\overline{(q^c_{L})^a_{i\alpha}}\; q^b_{Lj\beta}\, X^\kappa_m + h.c. \,,
\label{lag4}
\eea
where the new indices $\alpha,\beta=1,2$ are the $SU(2)_L$ indices,
$\tau^\kappa$ are the three Pauli matrices and $\varepsilon_{\alpha\beta}$
is the $2\times 2$ anti-symmetric tensor acting in the $SU(2)_L$ space.
In the Lagrangians above,
for the $SU(2)_L$ singlet diquarks the coupling matrices $f$ are
anti-symmetric in the generation space while for the $SU(2)_L$
triplets $f$ is symmetric. It follows that both singlets and triplets
can mediate the $t$-channel processes at hadron colliders while 
the $s$-channel processes can be induced only by the $SU(2)_L$ triplets.

We have calculated the cross sections of the above mentioned
processes mediated by
the diquarks at  Tevatron by convoluting over the default parton structure 
functions MRS (G) \cite{parton}
of the CERN Library package PDFLIB  \cite{pdflib} version 7.09.
The colour factors $F_c$  
coming from averaging over the initial state colours
and summing over the final state colours depend on the  $SU(3)_c$
representation. For the $s$- and $t$-channel cross sections 
we obtain $F_c=4/3$ and $F_c=2/3,$ and for the decay widths 
$F_c=2$ and $F_c=1$ corresponding to triplets and
sextets, respectively. 
For definiteness we shall in the following consider  the colour
sextet diquarks; 
the corresponding cross sections for the colour triplets 
are {\it larger} due to the larger colour factors.

Let us first study the $s$-channel resonance production of dijets
at Tevatron.
Neglecting the final state  fermion masses the diquark partial 
width is given by 
\bea
\Gamma=\frac{f^2}{8\pi}\,F_c\,M_X\,,
\label{gam}
\eea
which implies that the expected resonance is quite narrow.
Assuming that the charge 4/3 and -2/3 components of (6,3,1/3) diquark  
decay 100\% to $uu$ and $bb$ final states, respectively, we plot in Fig. 3
the dijet $s$-channel cross sections as functions of the diquark masses
for two values of collision energy $\sqrt{s}=1.8$ TeV and 
$\sqrt{s}=2$ TeV.
The diquark couplings are taken to be equal to unity.
One can see from the behaviour of the cross section 
curves in Fig. 3 that the narrow width approximation
is effective up to diquark masses $\sim 700$ GeV.
For considerably higher masses the width becomes larger according to
\Eq{gam} and 
the resonance peak will be smeared.
The difference in $p\bar p\to uu$ and $p\bar p\to bb$ cross sections is 
only due to the difference in the  parton structure functions
for the $u$ and $d$ quarks.

The narrow resonances in dijet signal \cite{dijet},
and in the double b-tagged dijet signal separately \cite{bbdijet},  
have been searched for with Tevatron CDF detector.
Negative results have resulted in model-independent bounds on the 
narrow resonance cross sections. Therefore,
for the diquark couplings equal to unity the present CDF data
allowes us to exclude the  charge 4/3 component of the
(6,3,1/3) diquark in the mass range 
\bea
270 \mrm{GeV}\lsim M_X\lsim 500 \mrm{GeV}
\eea
and the charge -2/3 component  in the mass
range 
\bea
200 \mrm{GeV}\lsim M_X\lsim 570 \mrm{GeV},
\eea
provided they decay 100\% to $uu$ and $bb$ pairs, respectively.

Next we consider the top quark pair production at Tevatron. In Fig. 4
we plot the cross sections of (6,1,4/3) or (6,3,1/3) diquark mediated
$t$-channel process $p\bar p\to t\bar t$
and the (6,3,1/3) diquark mediated $s$-channel 
resonance process $p\bar p\to t t$ as functions of the
diquark masses for two values of collision energy, as indicated in the figure.
The couplings $f$ are taken to be equal to unity. For the resonance
production we have assumed diquark branching ratio to $tt$
pair to be 100\%. With these assumptions the resonance cross
section exceeds the $t$-channel cross section for diquark masses 
below $\sim$ 700 GeV. Above that scale $t$-channel production becomes 
dominant. 

To derive constraints on the diquark masses and couplings we proceed as 
follows. The standard-model NLO (next to leading order) 
$t\bar t$ \cite{tt} and also $t\bar b$ \cite{tb} cross sections at the 
Tevatron and LHC are known within a total error of 15\%.
If the total  cross sections of the new processes exceed 15\% of the 
standard-model cross section, then the new signal will be detectable 
(such a criterion has recently been used in \cite{tait1}). 
Assuming that one of the cross sections dominates we find from 
the Tevatron Run I ($\sqrt{s}=1.8$ TeV, L=0.1 fb$^{-1}$)
$tt$ and $t\bar t$ cross sections that
the masses of  (6,3,1/3) and (6,1,4/3) diquarks
 should exceed 700 and 600 GeV, respectively. 
At Run II 
 ($\sqrt{s}=2$ TeV, L=2 fb$^{-1}$) the corresponding bounds would be 
750 and 600 GeV, respectively.

These bounds based on the cross section estimates indicate the sensitivity
of Tevatron to considered processes. 
Of course, dedicated Monte Carlo studies with appropriate kinematical 
cuts would allow one to achieve a much better   
signal-over-background ratio and therefore higher bounds on the diquark 
masses ( in particular at Run III, $\sqrt{s}=2$ TeV, L=30 fb$^{-1}$). 
This is because the diquarks are scalars while the 
standard model top pair production background  is dominantly produced 
in gluon-gluon and gluon-quark collisions. This statement 
applies especially to the $s$-channel $tt$ production. 
This  process is quark flavour violating and
the distribution of the final state top quarks is flat.
The unambiguous diquark signal in this process would be the same-sign
dilepton originating from top decays. This 
 has very little background from the standard model.
However, while large excess of the same-sign dileptons would indicate 
a clear signal of new physics then reconstruction of diquark
resonance peak in this channel is impossible due to the missing energy
carried away by neutrinos. On the other hand, studying the kinematics
of the visible particles in terms of endpoint spectra it should be
possible to extract information on the diquark mass also in this channel
(the Jacobian peak should be visible). This information can further be
combined with the studies of  hadronic and semileptonic channels.
These kind of dedicated studies are beyond the scope of the present
work.

To demonstrate the effectiveness of dedicated studies we first plot 
in Fig. 5 the cross section of the process $p\bar p \to t\bar c$
and $p\bar p \to t c$
against the mass of the (6,3,1/3) diquark $M_X$ for $\sqrt{s}=1.8$ TeV and 
$\sqrt{s}=2$ TeV assuming the couplings to be of order unity.
The single-top production at Tevatron due to anomalous chromomagnetic
couplings has been studied in \cite{tait2,han1}.
Assuming that  in the case of the diquark mediated process 
the same sensitivity to the $p\bar p \to t\bar c$ signal can be achieved as
in Ref.\cite{han1}, then comparison of the $p\bar p \to t\bar c$  
cross sections gives for Run I a bound $M_X\gsim 750$ 
GeV and  for Run II and III $M_X\gsim 1.2$ TeV and $M_X\gsim 1.6$ TeV, 
respectively.  These estimates are indicative of the sensitivity of the 
Tevatron to flavor-changing diquark processes.

For completeness, we obtain from the $t\bar b,$ $tb$ cross section estimates
at Tevatron that at Run I that the 
(6,3,1/3) diquark  mass should 
exceed 650 GeV and at Run II 720 GeV.   
Appropriate kinematical cuts and tagging of hard $b$ jets 
should separate the signal from the background.

Because  LHC will be a $pp$ collider, it will be the ideal place to 
search for scalar diquarks.  The most interesting process,
$pp\to tt,$ can be mediated by the $s$-channel (6,3,1/3) diquark.
Therefore, resonance production is possible. 
The  discussion on detectability of this process at Tevatron applies
also here. Another interesting process to study is  $pp\to bb$
which may also occur due to  the $s$-channel  (6,3,1/3) diquark
resonance. The resonance can be detected searching for doubly
$b$-tagged dijet final states. 
We plot the cross sections of these processes
in Fig. 6 as functions of $M_X.$ The couplings are taken equal to
unity and only one decay channel is assumed for the diquarks.
Considering $tt$ production,  the excess of 
same-sign charged dilepton final states will be
clear signal of the new physics. 
The main experimental error to this signal will
come from the misidentification of the lepton charge but this 
is expected to be small. To estimate which diquark mass scales 
can be probed at LHC we
assume that about $10^3$ events will constitute a discovery.
It follows from Fig. 6 that for  couplings of order unity, scalar diquarks
as heavy as 7 TeV can be discovered already at the low luminosity 
(L=10 fb$^{-1}$) run of  LHC. With the final luminosity,     
L=500 fb$^{-1}$, diquark masses of 13 TeV can be probed.
The same conclusions apply also to the process  $pp\to tc$
provided that the involved couplings are of order unity.

\subsection*{4. Conclusions}

We see from the above discussion that all exotic scalars which are bilinear 
combinations of two known fermion representations are all very heavy if their 
couplings are independent of the quark or lepton family and unsuppressed.
At closer scrutiny, we see also that 
the above derived bounds on the couplings and masses of the exotic scalars 
depend on the flavor indices and on the assumption of the existence of 
trilinear couplings.  If the latter couplings are absent, the most 
stringent bounds coming from the nonobservation of 
proton decay and neutron-antineutron oscillations, and from baryogenesis 
can be evaded in many cases.  Certain combinations of the couplings 
can be tested only in high-energy experiments.  The possibility exists for 
some of the diquark masses to be in the range of a few TeV, in which case 
they can be discovered at future hadron colliders. 
At Tevatron the best sensitivity to diquarks can be achieved by studying
the processes $\bar p p\to \bar t c, \bar t t.$ The resonance 
process $\bar p p\to  t t,$
though sea quark suppressed, may also provide an observable
excess of the same-sign dilepton final states indicating clearly for
the new physics. 
LHC will be an ideal facility for the studies of diquarks because 
the quark-flavor-violating resonance processes
$pp\to tt,bb$ will not be suppressed. Therefore diquark masses
as high as 13 TeV can be probed at LHC.

~\vskip 1.5in
\begin{center} {ACKNOWLEDGEMENT}
\end{center}

We would like to thank S. Ambrosanio, M. Diehl,
D. P. Roy, M. Spira and P.M. Zerwas 
for several discussions. Two of us
(MR and US) acknowledge financial support from the Alexander von 
Humboldt Foundation and hospitality of DESY Theory Group.
The work of EM was supported in part by the U.~S. Department of Energy 
under Grant No.~DE-FG03-94ER40837.

\newpage
\bibliographystyle{unsrt}

\begin{thebibliography}{99}

\bibitem{5} 
F. Cuypers and S. Davidson, Eur. Phys. J. {\bf C2}, 503 (1998).

\bibitem{leptoq} 
M. Leurer, \pr{D49}, 333 (1994);
S. Davidson, D. Bailey and B.A. Campbell, \zp{C61}, 613 (1994).


\bibitem{1} E. Ma and U. Sarkar, Phys. Rev. Lett. {\bf 80}, 5716 (1998).

\bibitem{volkas} 
J.P. Bowes, R. Foot  and R.R. Volkas, \pr{D54}, 6936 (1996). 

\bibitem{4} Particle Data Group, C. Caso {\it et al.}, Eur. Phys. J. 
{\bf C3}, 1 (1998).

\bibitem{3} A. Yu. Smirnov and M. Tanimoto, Phys. Rev. {\bf D55}, 1665 (1997).

\bibitem{2} For a recent overview, see for example E. Ma, Phys. Rev. Lett. 
{\bf 81}, 1171 (1998).

\bibitem{lepto} E.W. Kolb and M.S. Turner, {\it The Early Universe} 
  (Addison-Wesley, Reading, MA, 1989); V.A. Kuzmin, V.A Rubakov and 
  M.E. Shaposhnikov, Phys. Lett. {\bf B 155}, 36 (1985); J.A. Harvey 
  and M.S. Turner, Phys. Rev. {\bf D 42}, 3344 (1990); M. Fukugita 
  and T. Yanagida, Phys. Rev. {\bf D 42}, 1285 (1990);  
  S.M. Barr and A.E. Nelson,  Phys. Lett. {\bf B 246}, 141 (1991). 
\bibitem{RS} M. Raidal and A. Santamaria, \plet{B421}, 250 (1998).
\bibitem{double} T. Rizzo, \pr{D25}, 1355 (1982); \pr{D27}, 657 (1983); M. 
Lusignoli and S. Petrarca, \plet{B226}, 397 (1989); M.D. Swartz, \pr{D40},
1521 (1989); J.A. Grifols, A. Mendez and G.A. Schuler, Mod. Phys. Lett. 
{\bf A4}, 1485 (1989); J.F. Gunion, J.A. Grifols, A. Mendez, B. Kayser and 
F. Olness, \pr{D40}, 1546 (1989); J. Maalampi, A. Pietil\"a and M. Raidal, 
\pr{D48}, 4467 (1993); N. Lepore \ea, \pr{D50}, 2031 (1994); E. Accomando 
and S. Petrarca, \plet{B323}, 212 (1994); J.F. Gunion, \ijmp{A11}, 1551 
(1996); J.F. Gunion, C. Loomis and K.T. Pitts, hep-ph/9610327; K. Huitu, 
J. Maalampi, A. Pietil\"a and M. Raidal, \np{B487} (1997) 27; G. Barenboim, 
K. Huitu, J. Maalampi and M. Raidal, \plet{B394}, 132 (1997); F. Cuypers 
and M. Raidal, \np{B501}, 3 (1997); S. Chakrabarti, D. Choudhuri, R.M. 
Godbole and B. Mukhopadhyaya, hep-ph/9804297.

\bibitem{R} M. Raidal, \pr{D57}, 2013 (1998).

\bibitem{9} See for example E. Keith and E. Ma, Phys. Rev. Lett. {\bf 79}, 
4318 (1997).

\bibitem{10} H1 Collaboration, C. Adloff {\it et al.}, Z. Phys. {\bf C74}, 191 
(1997); ZEUS Collaboration, J. Breitweg {\it et al.}, Z. Phys. {\bf C74}, 207 
(1997).

\bibitem{11} M. Baldo-Ceolin {\it et al.}, Z. Phys. {\bf C63}, 409 (1994).

\bibitem{fc} T. P. Cheng and M. Sher, Phys. Rev. {\bf D35}, 3484 (1987);
{\bf D44}, 1461 (1991); A. Antaramian, L. J. Hall and A. Rasin, 
Phys. Rev. Lett. {\bf 69}, 1871 (1992);
L. J. Hall and S. Weinberg, Phys. Rev. {\bf D48}, R979 (1993).

\bibitem{eilam} G. Eilam, J. L. Hewett and A. Soni, \pr{D44}, 1473 (1991). 

\bibitem{parton} A.D. Martin, R.G. Roberts and W.J. Stirling,
RAL preprint, RAL/95-021 (1995).

\bibitem{pdflib} H. Plothow-Besch, Comp. Phys. Comm. {\bf 75}, 396 (1993);
Int. J. Mod. Phys. {\bf A10}, 2901 (1995).

\bibitem{dijet} CDF Collaboration, F. Abe  {\it et al.}, \pr{D55}, 5263 (1997).


\bibitem{bbdijet} CDF Collaboration, F. Abe  {\it et al.}, hep-ph/9809022.


\bibitem{tt} S. Catani, M. L. Mangano, P. Nason and L. Trentadue,
Phys. Lett. {\bf B 378}, 329 (1996);
E. Berger and H. Contopanagos, Phys. Rev. {\bf D54}, 3085 (1996).
\bibitem{tb} T. Stelzer, Z. Sullivan and S. Willenbrock, Phys. Rev.
{\bf D56}, 5919 (1997); and hep-ph/9807340.
\bibitem{tait1} T. Tait and C.-P. Yuan, hep-ph/9710372.
\bibitem{tait2} E. Malkawi and T. Tait, Phys. Rev. {\bf D54}, 5758 (1996).
\bibitem{han1} T.Han \ea, hep-ph/9806486.


\end{thebibliography}

\newpage
\begin{center}
\begin{picture}(300,150)(0,0)
\Photon(120,60)(180,60)34
\ArrowLine(80,0)(120,0)
\ArrowLine(220,0)(180,0)
\DashArrowLine(120,0)(180,0)3
\ArrowLine(120,100)(120,60)
\ArrowLine(120,60)(120,30)
\ArrowLine(120,30)(120,0)
\ArrowLine(180,100)(180,60)
\ArrowLine(180,60)(180,30)
\ArrowLine(180,30)(180,0)
\Text(100,-8)[c]{$u_R$}
\Text(200,-8)[c]{$l_R$}
\Text(150,-10)[c]{$(3^*,1,4/3)$}
\Text(115,80)[r]{$s_L$}
\Text(115,45)[r]{$c_L$}
\Text(115,15)[r]{$c_R$}
\Text(185,80)[l]{$u_L$}
\Text(185,45)[l]{$s_L$}
\Text(185,15)[l]{$s_R$}
\Text(150,68)[c]{$W$}
\end{picture}
\vskip 0.5in
{\bf Fig.~1.} ~ One-loop proton decay due to the $(3^*,1,4/3)$ scalar.

\begin{picture}(300,150)(0,0)
\ArrowLine(70,0)(110,0)
\ArrowLine(190,0)(110,0)
\ArrowLine(190,0)(230,0)
\Text(90,-8)[c]{$u_L(d_L)$}
\Text(150,-8)[c]{$d_L(u_L)$}
\Text(210,-8)[c]{$\nu_L(l_L)$}
\DashArrowLine(110,0)(150,40)3
\DashArrowLine(190,0)(150,40)3
\DashArrowLine(150,80)(150,40)3
\Text(125,20)[r]{$(3^*,1,1/3)$}
\Text(176,20)[l]{$(3^*,1,1/3)$}
\Text(155,60)[l]{$(3^*,1,-2/3)$}
\ArrowLine(122,108)(150,80)
\ArrowLine(178,108)(150,80)
\Text(120,114)[c]{$d_R$}
\Text(180,113)[c]{$s_R$}
\end{picture}
\vskip 0.5in
{\bf Fig.~2.} ~ One-loop proton decay due to the $(3^*,1,-2/3)$ scalar.
\end{center}

\newpage

\begin{figure}[htb]
\centerline{
\epsfxsize = 0.9\textwidth \epsffile{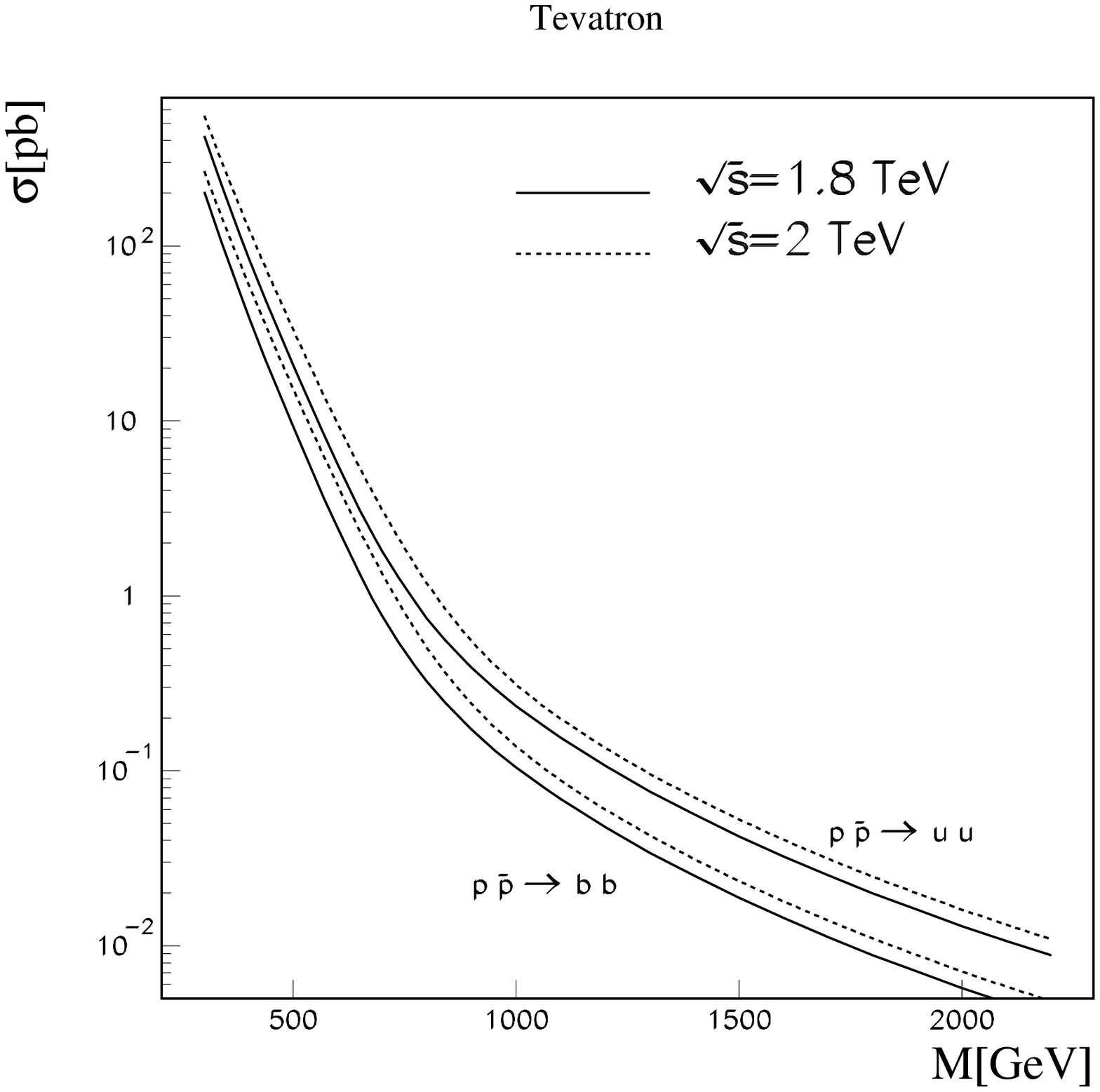} 
}
\end{figure}

{\bf Fig.~3.} ~ Cross sections of the $s$-channel processes $p\bar p\to uu$
and $p\bar p\to bb$
at Tevatron as functions of the scalar (6,3,1/3) 
diquark masses. The couplings are 
taken to be equal to unity. 

\newpage

\begin{figure}[htb]
\centerline{
\epsfxsize = 0.9\textwidth \epsffile{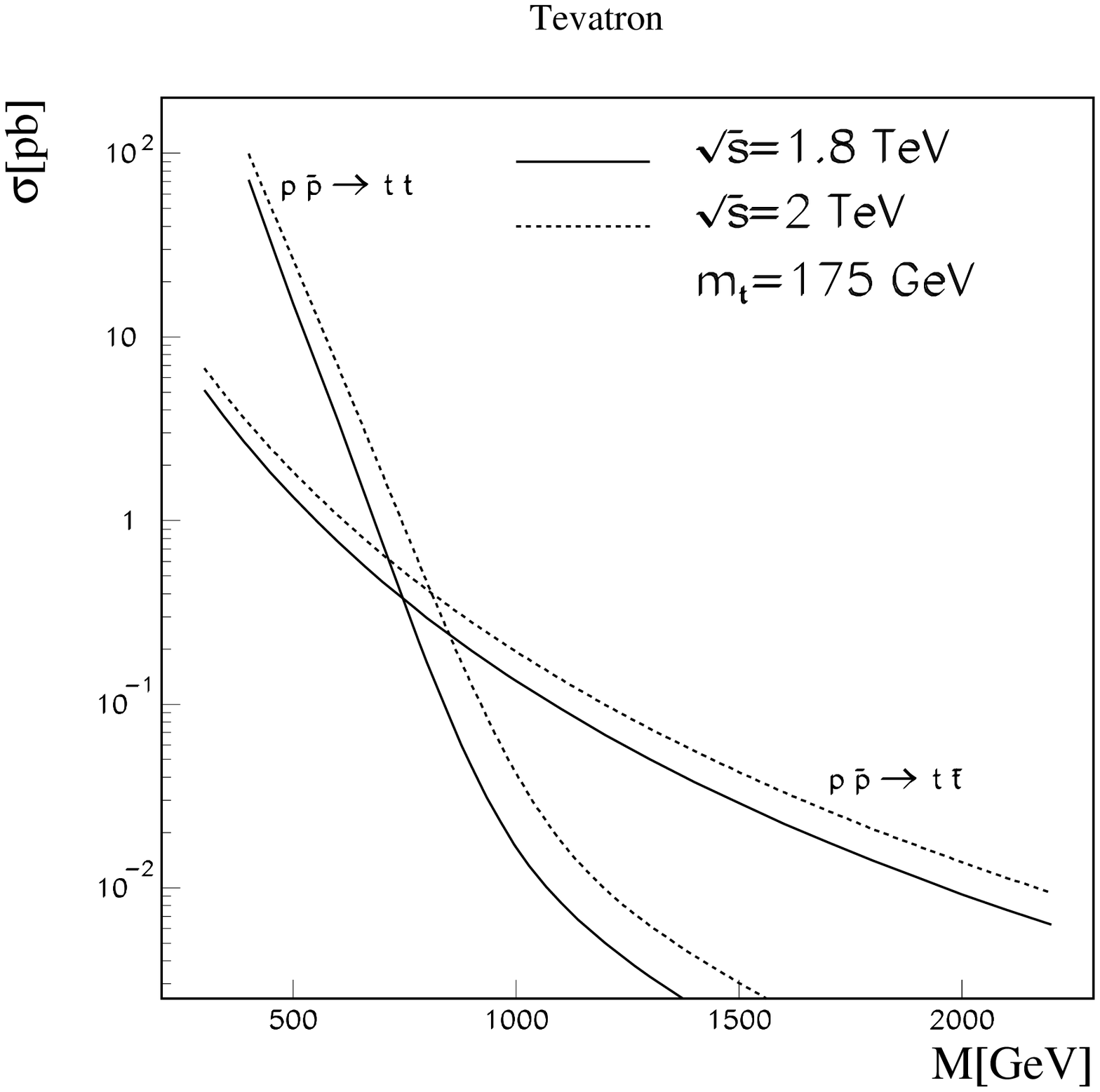} 
}
\end{figure}

{\bf Fig.~4.} ~ Cross sections of the processes $p\bar p\to t\bar t$ and 
$p\bar p\to t t$ at Tevatron against the (6,3,1/3) scalar diquark 
mass $M_X$. The couplings are taken to be equal to unity. 

\newpage

\begin{figure}[htb]
\centerline{
\epsfxsize = 0.9\textwidth \epsffile{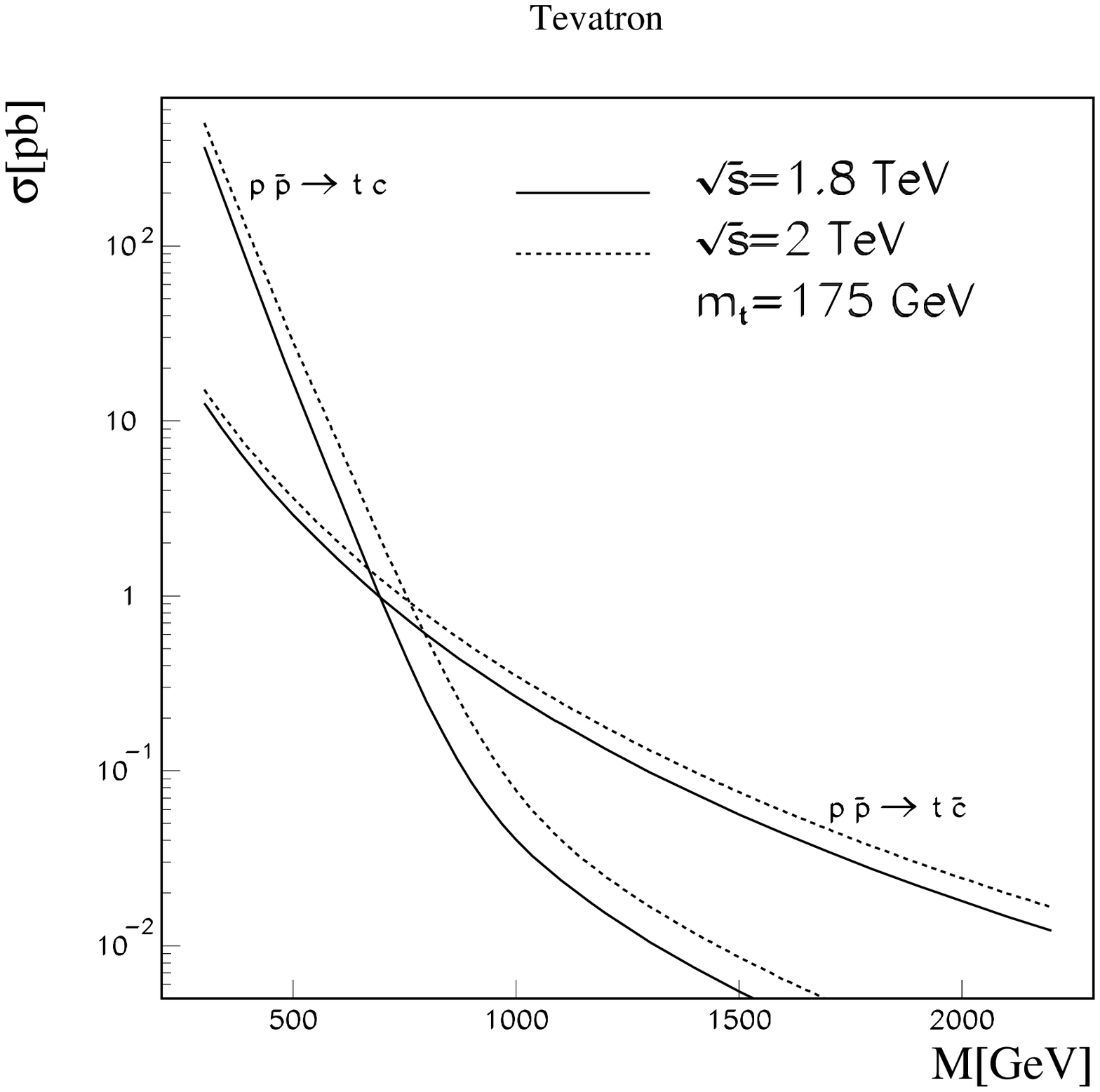} 
}
\end{figure}

{\bf Fig.~5.} ~ Cross sections of the FCNC processes $p\bar p\to t\bar c$
and $p\bar p\to t c$
at Tevatron against the (6,3,1/3) scalar diquark mass $M_X$. 
The couplings are 
taken to be equal to unity. 

\newpage

\begin{figure}[htb]
\centerline{
\epsfxsize = 0.9\textwidth \epsffile{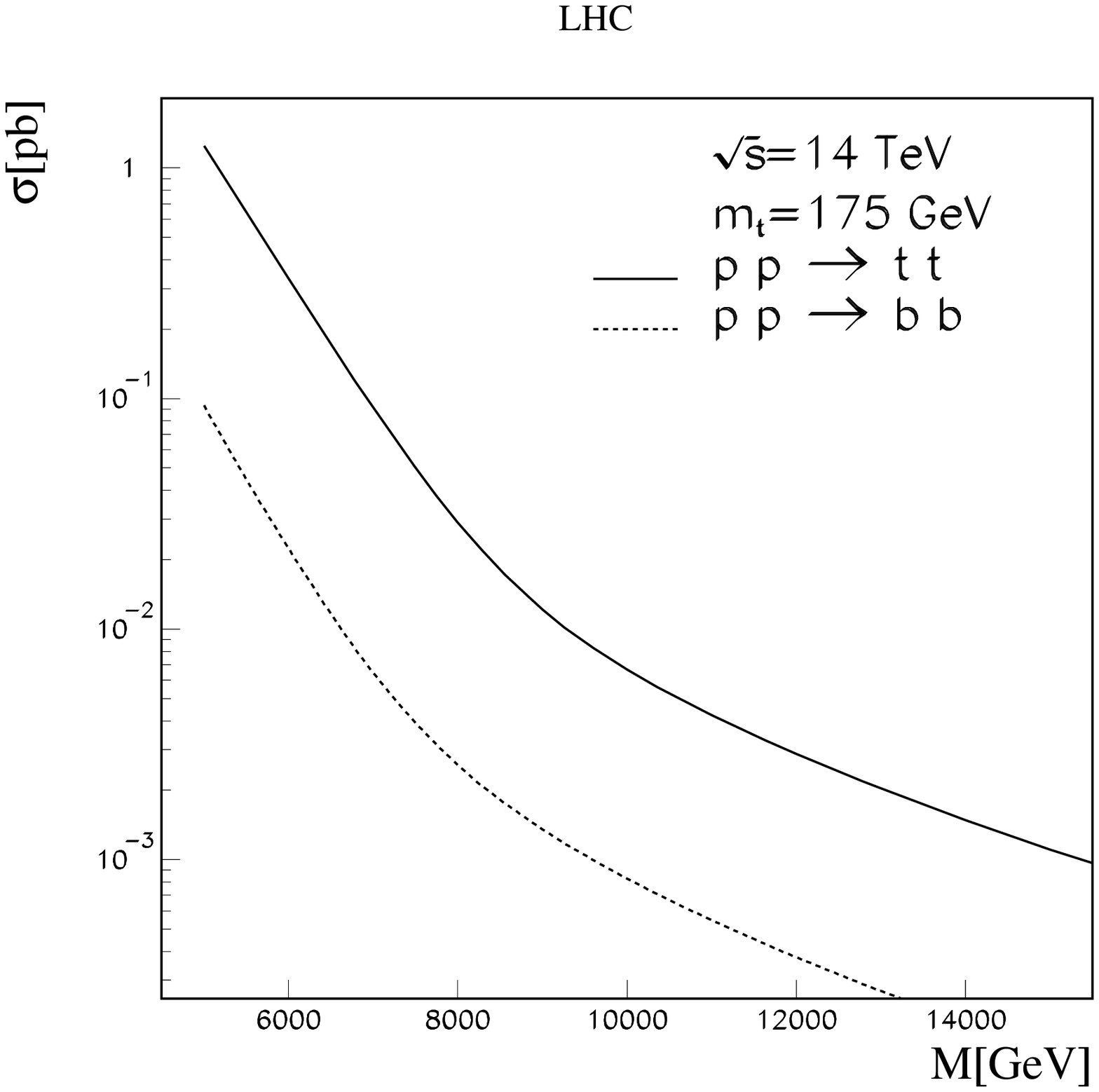} 
}
\end{figure}

{\bf Fig.~6.} ~ Cross sections of the $s$-channel 
process $p p\to tt$ and $p p\to bb$
at LHC against the  scalar (6,3,1/3) diquark masses. The couplings are 
taken to be equal to unity. 

\end{document}